\pdfoutput=1
\documentclass[aps,pre, twocolumn, groupedaddress]{revtex4-1}
\usepackage{lipsum}
\usepackage{mathtools}
\usepackage{graphicx}
\usepackage{dcolumn}
\usepackage{amsmath}    
\usepackage{amssymb}
\usepackage{bm}
\usepackage{hyperref}
\usepackage{latexsym}
\usepackage{verbatim}
\usepackage{color}
\usepackage{float}
\usepackage[caption=false]{subfig}
\usepackage{epstopdf}

\def\beq{\begin{equation}}
\def\eeq{\end{equation}}

\def\beq{\begin{equation}}                          
\def\eeq{\end{equation}}                          
\def\bea{\begin{eqnarray}}                          
\def\eea{\end{eqnarray}}

\DeclareRobustCommand{\uvec}[1]{{%
  \ifcsname uvec#1\endcsname
     \csname uvec#1\endcsname
   \else
    \bm{\hat{\mathbf{#1}}}%
   \fi
}}

\preprint{}

\begin{document}

\title{Effect of polydispersity on the dynamics of active Brownian particles}

\author{Sameer Kumar}
\email[]{sameerk.rs.phy16@itbhu.ac.in}
\affiliation{Department of Physics, Indian Institute of Technology (BHU), Varanasi, U.P. India - 221005}

\author{Jay Prakash Singh}
\email[]{jayp.rs.phy16@itbhu.ac.in}
\affiliation{Department of Physics, Indian Institute of Technology (BHU), Varanasi, U.P. India - 221005}

\author{Debaprasad Giri}
\email[]{dgiri.app@iitbhu.ac.in}
\affiliation{Department of Physics, Indian Institute of Technology (BHU), Varanasi, U.P. India - 221005} 

\author{Shradha Mishra}
\email[]{smishra.phy@itbhu.ac.in}
\affiliation{Department of Physics, Indian Institute of Technology (BHU), Varanasi, U.P. India - 221005}

\begin{abstract}
We numerically study the dynamics and the phases of self-propelled disk-shaped particles of different sizes with soft repulsive potential in two dimensions.  Size diversity is introduced by the polydispersity index (PDI) $\epsilon$, which is the width of the uniform distribution of the particle's radius. The self-propulsion speed of the particles controls the activity $v$. We observe enhanced dynamics for large size diversity among the particles. We calculate the effective diffusion coefficient $D_{eff}$ in the steady-state.
The system exhibits four distinct phases, jammed phase with small $D_{eff}$ for small activity and liquid phase with enhanced $D_{eff}$ for large activity. The number fluctuation is larger and smaller than the equilibrium limit in the liquid and jammed phase, respectively.
Further, the jammed phase is of two types: solid-jammed and liquid jammed for small and large PDI. Whereas the liquid phase is called motility induced phase separation (MIPS)-liquid for small PDI and for large PDI, we find enhanced diffusivity and call it the {\em pure liquid} phase. The system is studied for three packing densities $\phi$, and the response of the system for polydispersity is the same for all $\phi$'s. Our study can help understand the behavior of cells of various sizes in a tissue,  artificial self-driven granular particles, or living organisms of different sizes in a dense environment.

\end{abstract}

\maketitle
\section{Introduction}
The dynamics of self-propelled particles perpetually moving by converting energy from the environment into mechanical motion and collisions represent a non-equilibrium phenomenon. Such non-equilibrium systems exhibit many interesting properties such as clustering, collective motion [\cite{GopinathPRE2012, PeruaniPRE2006}], anomalous density fluctuations [\cite{RamaswamySimhaToner2003}], strange rheological behavior [\cite{GiomiPRE2010,  SaintillanPRE2010, CatesPRL2008}], and activity-dependent phase change [\cite{ShenPNAS2004}].
Their size ranges from few microns, e.g., bacteria [\cite{DombrowskiPRL2004}, cells [\cite{KemkemerEPJE2000}], cytoskeletal filament [\cite{SurreyScience2001}], motor proteins [\cite{BendixBPJ2008}], etc., to macroscopic systems like fish school, birds flock, and animal herds [\cite{VicsekPR2012}], etc.\\
In 1995 Vicsek and coworkers [\cite{VicsekPRL1995}] proposed a swarming model, one of the building blocks to study active matter systems [\cite{SameerPRE2020, SinghPhysicaA2020}].   
Colloidal Janus particles [\cite{ JiangPRL2010, VolpeSM2011}], which act as an artificial microswimmer due to its asymmetry of surface chemistry, were considered the model system for active matter, often called the active particles.  Active particles are generally of two types based on their appearance;  elongated rod-like particles are called polar/apolar particles [\cite{MarchettiRMP2013}], and spherically symmetric particles fall in the category of the active Brownian particles (ABPs). These micron-sized ABPs move in an environment with a low Reynolds number, and hence their dynamics, in general, are overdamped [\cite{HowsePRL2007, HagenCMP2009, HagenJPhysCM2011, HagenPRL2013}]. The active Brownian motion appears due to the interplay of self-propulsion and the thermal noise in the system and verified experimentally by studying the collective behavior of colloids and bacteria [\cite{HowsePRL2007, HagenCMP2009, HagenPRL2013, KurzthalerPRL2018, BechingerRMP2016}].\\
Recent studies address the dynamics of ABPs on various environmental backgrounds, e.g., the motion of ABPs on a periodic substrate, channel-based transport of ABPs [\cite{PattanayakEPJE2019, PabloSM2021}], 
and dynamics of ABPs in a confined geometry 
[\cite{SDasPRE2020,  MishraPhysicaA2017, ReversatNature2020}], etc. In these studies, 
apart from the different nature of particle-to-particle interaction  
(for example, hard or soft repulsive interaction) 
[\cite{ DolaiSM2018}], particles are, in general, considered to be of the 
same size, i.e., monodisperse. But, there are many cellular systems, 
bacteria, and colloids that possess size diversity, 
i.e., all particles do not necessarily have the same radius and 
can be termed as polydisperse. 
The polydispersity of the particles' size can lead to many interesting properties in terms of their dynamics.  \\
The self-propelled particle (SPP) model has been described in 
[\cite{ BelmontePRL2008, GarciaPNAS2015, HenkesPRE2011, SepulvedaPLoS2013, SoumyaPLOS2015, SzaboPRE2006}] to study such systems. 
These models are similar to those for inert particulate matter where cells (or bacteria etc.) are represented as disks or spheres that interact with an isotropic soft repulsive potential and electrostatic attraction.  SPP models typically exhibit a glass transition from a 
diffusive fluid state to an arrested subdiffusive solid that is controlled by (1) the strength of self-propulsion [\cite{GarciaPNAS2015,  HenkesPRE2011, NiNatCom2013}]
and (2) the packing fraction $\phi$ [\cite{HenkesPRE2011, NiNatCom2013,  BerthierPRL2014,  FilySM2014, FilyPRL2012}]. Polydispersity also plays a crucial role in these transitions, 
and it is important to study the effect of particles' size diversity on the steady-state phase of the system. {The effect of polydispersity have been seen in the equilibrium systems [\cite{BommineniPRL2019, NiCommPhys2019, NiJCP2020}], but the understanding is very limited in the non-equilibrium counterparts.}
In, [\cite{HenkesPRE2011}] authors have considered self-propelled particles dynamics with some polydispersity, but they do not
explicitly explain the effect of particles' size diversity in the system dynamics. In [\cite{YethirajPRL2012}], authors have studied the dynamics of tracers with quenched polydispersed obstacles where they have addressed the system for a different amount of polydispersity in the obstacles' size and its effect on the percolation density. \\
In this study, we address the consequence of polydispersity and activity on the dynamics of ABPs. We use overdamped Langevin's dynamics to study the particles' motion in two dimensions.  The polydispersity index, $ \epsilon $,  characterize the diversity in the particles' size, which is the width of a uniform probability distribution of particles' radius. In contrast, the self-propulsion speed of the particles characterize the activity. Also, the system is studied for three different packing densities $\phi=\lbrace 0.65, \ 0.75, \ 0.85 \rbrace$. We do not exceed $\phi=0.85$ as the cut-off packing fraction remains under the shape rigidity limit is $0.85$, [\cite{BoltonPRL1990}]. We calculate the steady-state diffusion coefficient $D_{eff}$ and for large activity $v$, it follows a scaling function $D_{eff} \sim D_0 v^{\beta} f(\epsilon v^{-\alpha})$, where $\alpha$ and $\beta$ are the two exponents. We find system exhibits {\em four} distinct phases. The system is in the jammed and liquid phase for small and large activities. The jammed phase is characterized by small $D_{eff}$ and, it is of two types: solid jammed for small  PDI and liquid jammed for large PDI. The liquid phase is again of two types: MIPS-liquid [\cite{FilyPRL2012, RednerPRL2013, StenhammarPRL2013, TailleurPRL2008,  GonnellaIJMPC2014, SumaEPL2014,  BerthierPRL2014, WittkowskiNatComm2014, CatesARCMP2015, ZhanATS2020}] for small PDI with moderate $D_{eff}$ and {\em pure liquid} phase with {\em enhanced diffusivity}. The number fluctuation  larger [\cite{RamaswamySimhaToner2003, TonerAoP2005, TonerPRL1995}] and smaller [\cite{ HenkesPRE2011}] than the equilibrium limit in liquid and jammed phase respectively. \\
We divide the rest of the article in the following manner. In section \ref{model400}, we discuss the model used to study the system;  in section \ref{results400} we discuss about the results and finally summarise in section \ref{discussion400}.

\begin{figure}      
\centerline{\includegraphics[width=\linewidth]{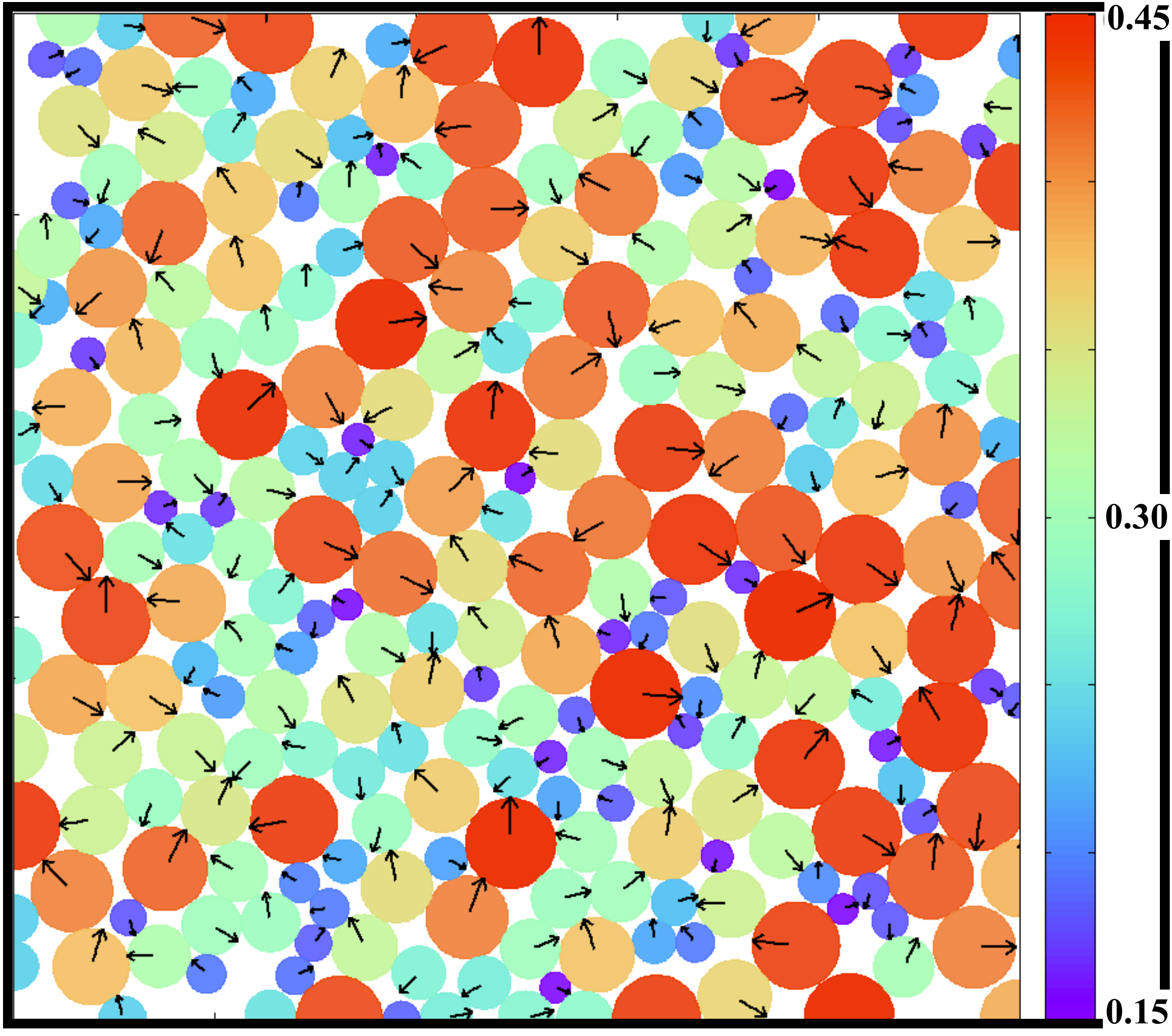}} 
\caption{ (Color online) Snapshot of the system for a non-zero polydispersity i.e. $\epsilon = 0.5$. Disks resembles the active Brownian particles with their radius represented by color bar,  and the arrows shows their velocity direction. }
\label{fig: 401}
\end{figure}

\section{Model and numerical details} 
\label{model400}
We distribute the particles randomly on a two-dimensional substrate. Radius, $R_i$, of particles is taken from a uniform distribution $P(R_i, \epsilon_0) \in [R_0-\frac{\epsilon_0}{2}, \ R_0+\frac{\epsilon_0}{2}]$, where $R_0$ 
is the mean radius and $\epsilon_0$ is the width of the distribution. 
We use over-damped Langevin's dynamics to study the particles motion which is given as, 
\begin{equation}
\centerline{$ \frac{\partial}{\partial t}{\bf r}_i(t)= v_0 {\bf \hat{e}}_i + \mu \sum_j^N {\bf F}_{ij}$}	
\label{eq: 401}
\end{equation}
\begin{equation}
\centerline{$ \frac{\partial}{\partial t}{\bf \theta }_i(t)= \sqrt{2D_R} \eta_i^R(t)$}  
\label{eq: 402}
\end{equation}
Here, ${\bf r}_i(t)$ is the position of $i^{th}$ particle  at  time $t$, $v_0$ is the self-propulsion speed which is same for all the particles and, $\theta_i(t)$ is the orientation angle which defines $ {\bf \hat{e}}=(cos (\theta), sin (\theta))$.  The interaction force between the particles is, ${\bf F}_{ij}=-\nabla U(r_{ij})$, where $U(r_{ij})$ is a harmonic potential  defined as,
\begin{equation}
\centerline{$U(r_{ij})=\frac{\kappa}{2}(r_{ij}-\sigma_{ij})^2 \Theta(1-\frac{r_{ij}}{\sigma_{ij}})$}
\label{eq: 403}
\end{equation}
Here, $\Theta(x)=1$ for $x \geq 0 $ and; $\Theta(x)=0$ for $x < 0 $. $r_{ij}=|{\bf r_i - r_j}|$ is the separation between two particles and 
$\sigma_{ij}={R}_i+{R}_j$. $\kappa$ is the force  constant. {$\mu$ is the mobility and is 
inversely proportional to the friction coefficient such that each particle is driven by a constant force of magnitude 
equal to $\frac{v_0}{\mu}$. $(\mu \kappa)^{-1}$ is the elastic time scale}. {$\eta$ is the random Gaussian white noise 
with $\langle \eta({\bf r},t)\rangle=0 $ and $\langle  \eta({\bf r},t) \eta({\bf r'},t') \rangle= \delta({\bf r}-{\bf r'})\delta(t-t') $, 
here $D_R$ is the rotational diffusion coefficient. $D_R^{-1}$ is the time scale over which the orientation of an active particle changes. Hence, $l_p = v_0 D_R^{-1}$, the persistence length 
or run length, is the typical distance travelled by an active particle before it changes direction}.
We keep the mobility and rotational noise fixed throughout the whole study i.e.,  $\mu=1.0$ and $D_R=1.0$. 
Whereas, the system is studied for $v_0\in (0.1, \ 1.0)$,  $\epsilon_0 \in (0.0, \ 0.25)$. We study the system for three different packing densities,  $\phi= 0.65, 0.75$ and $0.85$, which is defined as $\phi=\frac{\Sigma_i^N \pi R_i^2}{L^2} $, where $L$ is the size of the system, and $N$ is the total number of particles. We keep the mean radius  fixed, i.e. $R_0=0.3$. We redefine dimensionless activity, $v = \frac{v_0}{ R_0 \mu \kappa}$ and the dimensionless polydispersity, $\epsilon=\frac{\epsilon_0}{R_0}$ which is termed as polydispersity index (PDI). \\
We simulate the system in a square box of $L \times L $ with periodic boundary conditions.{ We choose $L=20$ for most of the simulation data, otherwise mentioned}. We start with a random homogeneous distribution of the particles in the box and with random directions.  Fig. \ref{fig: 401}, shows the  snapshot of the system generated from the simulation for a  non-zero polydispersity, $ \epsilon =0.5$. The Center of the disks show their position, ${\bf r}$  in the xy-plane, and the arrow on it implies the velocity direction, ${\bf \hat{e}}$. Equations (\ref{eq: 401} - \ref{eq: 403}) are updated for all particles and one simulation step is counted after a single update for all the particles. The steady-state in the system is achieved after simulation time $\mathcal{O}(10^5)$, the maximum simulation time is $10^7$ and we take the step size for the times $\Delta t= 10^{-3}$.  We use $15$ independent realization for averaging the data. 
\begin{figure}      
\centerline{\includegraphics[width=\linewidth]{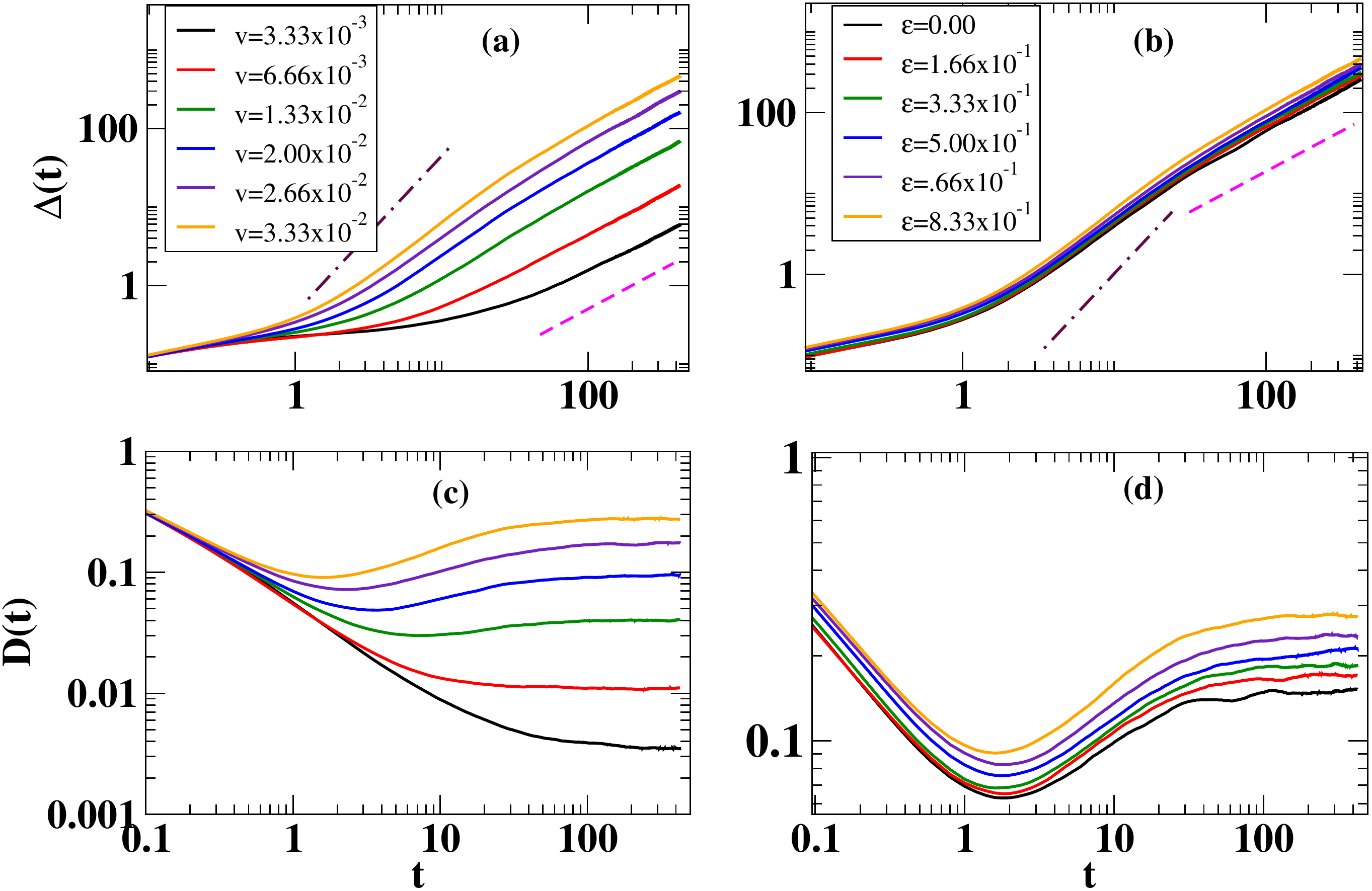} }
\caption{ (Color online) (a) Mean squared displacement, $\Delta (t) $ vs. $t$ for different actyivity ($v$) and fixed polydispersity index, $\epsilon=8.33 \times 10^{-1}$. (b) $\Delta (t) $ vs. $t$ for different $\epsilon$ and fixed activity, $v=3.33 \times 10^{-2}$. (c) Diffusion coefficient, $D(t)$ vs. $t$ for different actyivity ($v$) and fixed polydispersity index, $\epsilon=8.33 \times 10^{-1}$. (d)  $D(t)$ vs. $t$ for different $\epsilon$ and fixed activity, $v=3.33 \times 10^{-2}$.}
\label{fig: 402}
\end{figure} 
\begin{figure*}      
\includegraphics[width=\linewidth]{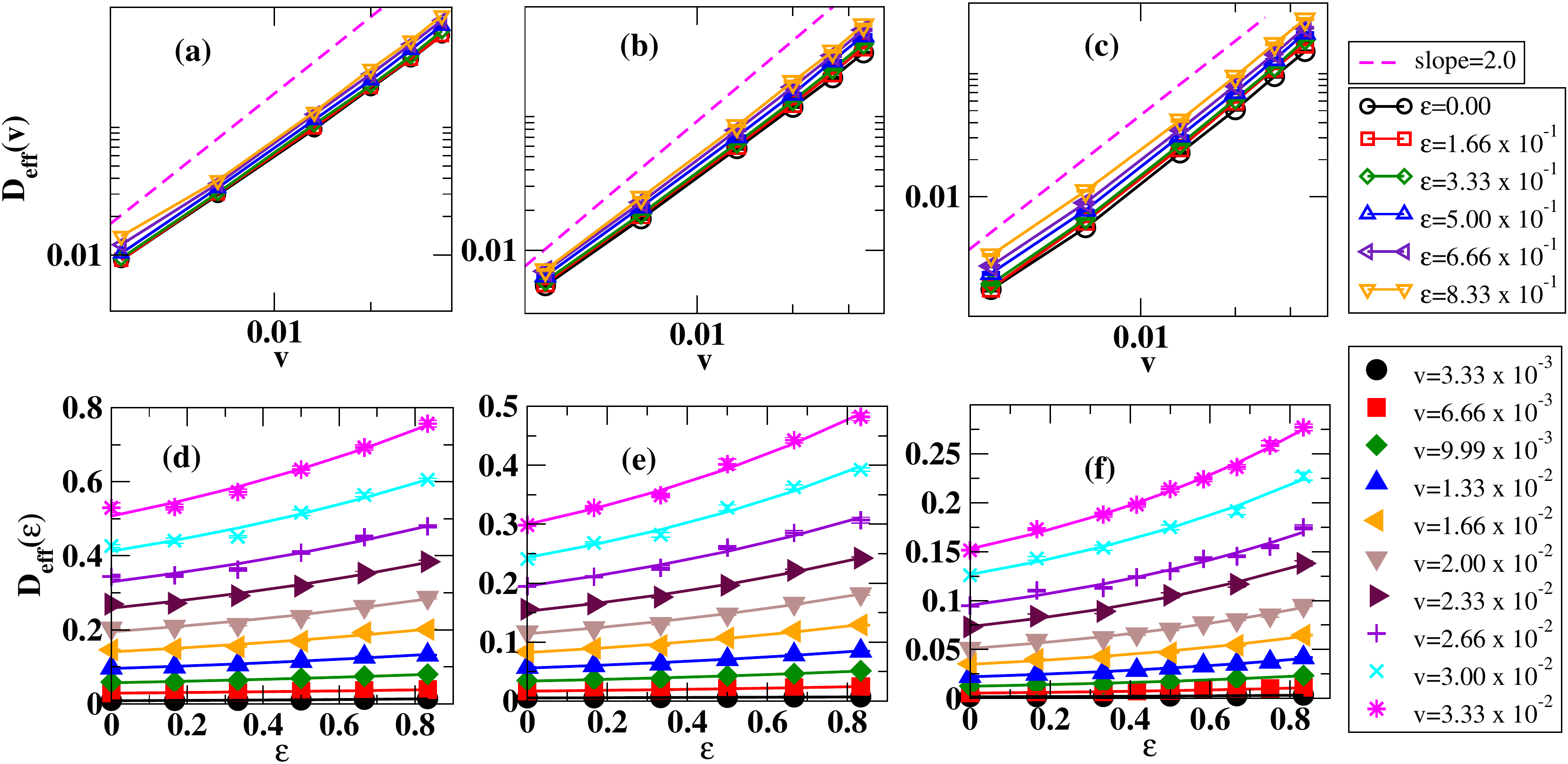} 
\caption{ (Color online) Effective diffusion coefficient, $D_{eff}$ vs. activity $v$ for different polydispersity index ($\epsilon$),  for $\phi=0.65$ (a), $0.75$ (b), $0.85$ (c). Effective diffusion coefficient $D_{eff}$ vs. polydispersity index, $\epsilon$ for different activity ($v$). Symbols are from the simulation and the solid lines are the fit (Eq. \ref{eq: 404}) to the data points. Different plots are for  different values of $\phi=0.65,0.75, 0.85$  (d-f), respectively. Error bars are smaller than the size of the symbols.}
\label{fig: 403}
\end{figure*} 

\section{Results}
\label{results400}
We calculate the different physical quantities and based on that we characterise the system properties under the different conditions for system variables i.e. packing density $\phi$, activity $v$ and polydispersity, $\epsilon$.
\subsection{Enhanced Diffusivity for finite Polydispersity}
\begin{figure*}      
\centerline{\includegraphics[width=\linewidth]{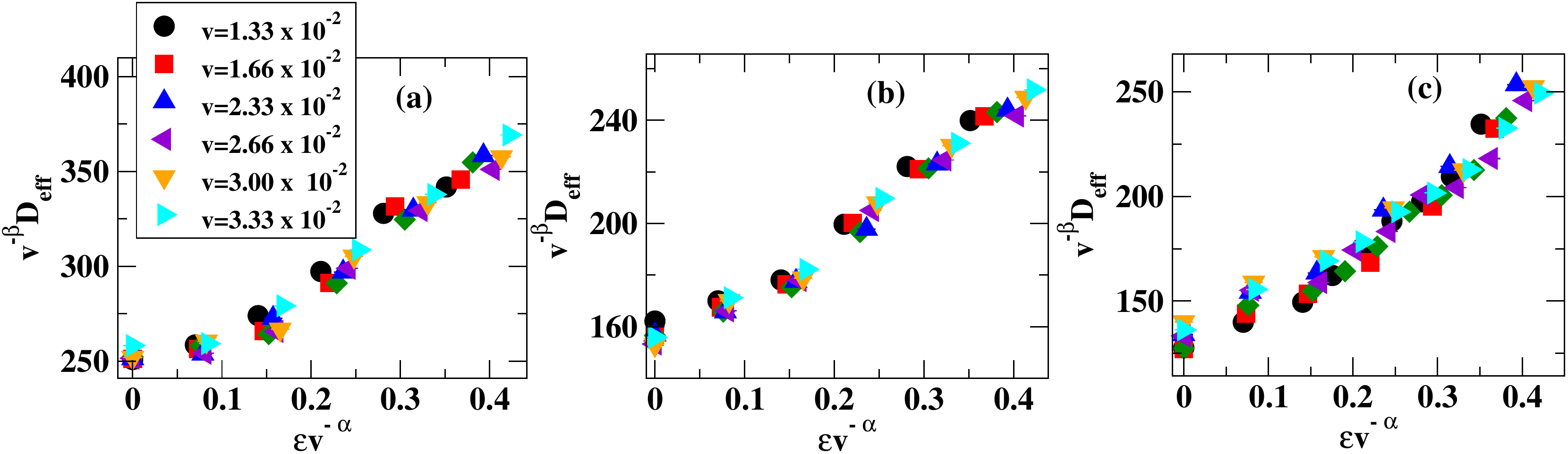}} 
\caption{ (Color online)  Scaled diffusivity, $D_{eff}/ v_0^{\beta}$ vs. scaled PDI, $\epsilon v^{-\alpha}$  for $\phi=0.65, \ 0.75, \ 0.85$ (a-c), respectively; where  $\alpha = -0.2$ and $\beta = 2.0$}
\label{fig: 404}
\end{figure*}

We characterize the system's properties by calculating the mean squared displacement of the particles for different sets of parameter. First, we calculate the mean squared displacement (MSD) $\Delta (t)$ defined as $\Delta(t)=\langle \frac{1}{N}\sum_i^N \vert {\bf r}_i(t)-{\bf r}_i(0) \vert^2 \rangle$, where $\langle.....\rangle$ means the average over many initial configurations. Fig. \ref{fig: 402}, shows the plot of MSD and diffusion coefficients, $D(t)= \frac{\Delta(t)}{4t}$, for different activity  $v$ and polydispersity  index $ \epsilon$ for packing density $\phi=0.85$. In fig. \ref{fig: 402}(a) we plot the MSD for different activity $v$ and fixed PDI, $\epsilon=8.33 \times 10^{-1}$. We perceive that the system shows essentially two regimes of dynamics; first, it is super-diffusion for early time ($t<100$) where the slope of $\Delta (t)$ vs $t$ is greater than $1.0$, then starts diffusing later ($t>100$), where the MSD grows linearly with time, i.e., the persistent random walk (PRW). The super diffusion regime is a signature of active systems. At the start of the simulation, we witness a plateaus region due to the over-damped dynamics.  Further, when we increase the value of $v$, $\Delta (t)$ shift upward in the positive y-axis i.e. increase in diffusion coefficient $D(t)$ (\ref{fig: 402}(c)). Next, in figure \ref{fig: 402}(b), we see a similar trend for a fixed activity and different  polydispersity, i.e., as we increase PDI ($\epsilon$), MSD shift upwards. This leads to the diffusion coefficient increase, shown in $D(t)$ vs. $t$ plot in fig. \ref{fig: 402}(d). We obtain similar pattern of changes in $\Delta (t)$ and $D(t)$ for ($v, \epsilon$) with packing density $\phi=0.65$ and $0.75$ (data not shown). \\
We do similar calculation for other parameters and plot the effective diffusion coefficient in fig. \ref{fig: 403} for different packing fractions. We define the effective diffusion coefficient $D_{eff}$ in steady state ($t>100$) as  $D_{eff} = \lim_{t \longrightarrow \infty} D(t)$ for different activity $v$ and PDI $\epsilon$. Fig. \ref{fig: 403}(a-c) shows the variation of $D_{eff}$ for different activity and for different packing fractions, $\phi$. We found that the effective diffusivity vs. activity have a slope $\beta \simeq  \ 2.0$ for  all $\phi's=0.65, \ 0.75 \ and \ 0.85$. Further, in fig. \ref{fig: 403}(d-f), we plot $D_{eff}$ vs $\epsilon$, and fit the data points with the  expression for diffusion coefficient fitted by, 

\begin{equation}
\centerline{$ D_{eff}(\epsilon,v)=v^{\beta} D_0(\phi)[1+exp({\frac{\epsilon}{\epsilon_c(v)}})]$}
\label{eq: 404}
\end{equation}

 where, $D_0(\phi)$ and $\epsilon_c(v)$, are the two fitting parameters depend on packing fraction and activity $v$, respectively. It shows that as we increase the size diversity among the ABPs, diffusivity of the system increases and the change is high for high activity in the system (which is explained in more detail in the next paragraph). In Fig. \ref{fig: 404} (a-c) we plot the scaled diffusivity $D_{eff}/v^{\beta}$ vs. scaled PDI, $\epsilon v^{-\alpha}$, for higher activity $v \geq 1.33 \times 10^{-2}$ and for three different packing densities $\phi=0.65, \ 0.75$ and $0.85$, respectively.  Interestingly we find a good collapse of data for all $\phi$'s and range of activities $v \geq 1.33 \times 10^{-2}$. The two exponents $\alpha$ and $\beta$ have values $-0.2$ and $2.0$, respectively. 
 The above scaling suggest the form of $D_{eff} \sim D_0 v^{\beta}f(\epsilon v^{-\alpha})$ and $f(x \rightarrow 0) \sim 2$, which is obtained from the proposed form for $D_{eff}$ in Eq. (\ref{eq: 404}).

\begin{figure*}      
\centerline{\includegraphics[width=\linewidth]{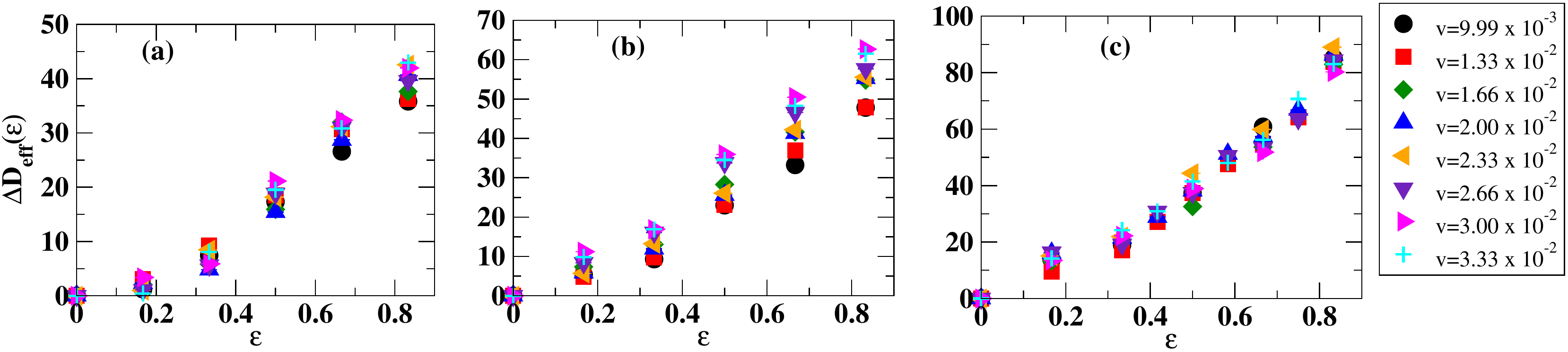} }
\caption{ (Color online) Percentage change in the effective diffusion coefficient, $\Delta D_{eff}$ vs. PDI, $\epsilon$ for different activity ($v$),  for $\phi=0.65$ (a), $0.75$ (b), $0.85$ (c).} 
\label{fig: 405}
\end{figure*}

{We also calculate the percentage change in the effective diffusion coefficient with respect to zero polydispersity i.e. $ \Delta D_{eff}(\epsilon)=\frac{D_{eff}(\epsilon)-D_{eff}(0)}{D_{eff}(0)} \times 100$. We plot $\Delta D_{eff}$ vs. $\epsilon$ as shown in fig. \ref{fig: 405}.  The value of $\Delta D_{eff}$ increases as we increase the packing density, as depicted from fig. \ref{fig: 405}(a, b and c) that show the percentage change in the effective diffusion coefficient for packing density, $\phi=0.65, \ 0.75$ and $0.85$, respectively. This feature indicates that the impact of polydispersity is more prominent in a dens system. Also, the change in $D_{eff}$ goes up to 100 percent for the largest polydispersity, see fig. \ref{fig: 405}(c).}
  
\begin{figure}      
\centerline{\includegraphics[width=\linewidth]{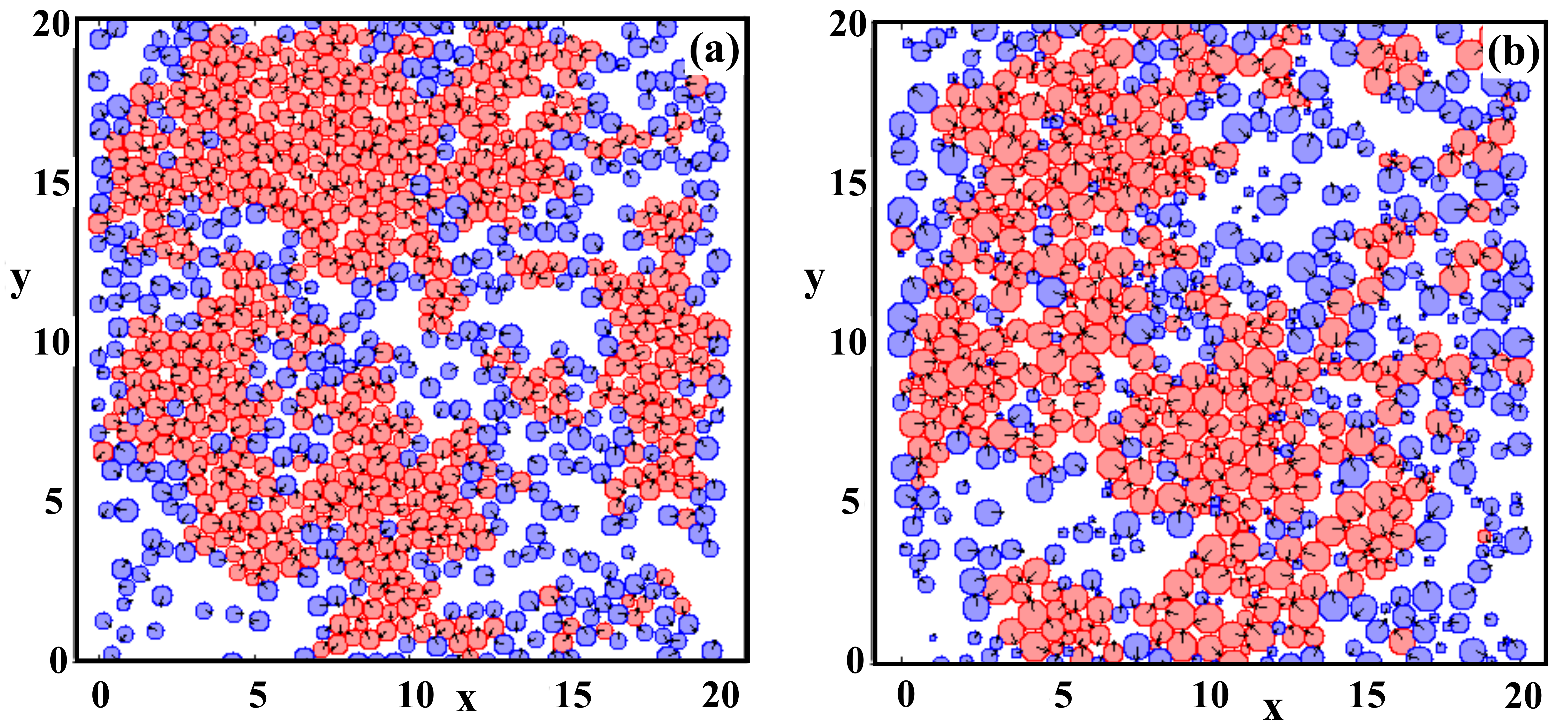}}
\caption{ (Color online) Snapshots at equal time for $\epsilon=1.66 \times 10^{-1}$(a), and $\epsilon=6.66 \times 10^{-1}$ (b), for fixed activity, $v=3.33 \times 10^{-2} $, and $\phi=0.65$. Rattlers are represented by blue 	disks whereas non-rattlers are in red. Arrows on the disks shows their velocity direction. } 
\label{fig: 406}
\end{figure} 

\begin{figure}      
\centerline{\includegraphics[width=\linewidth]{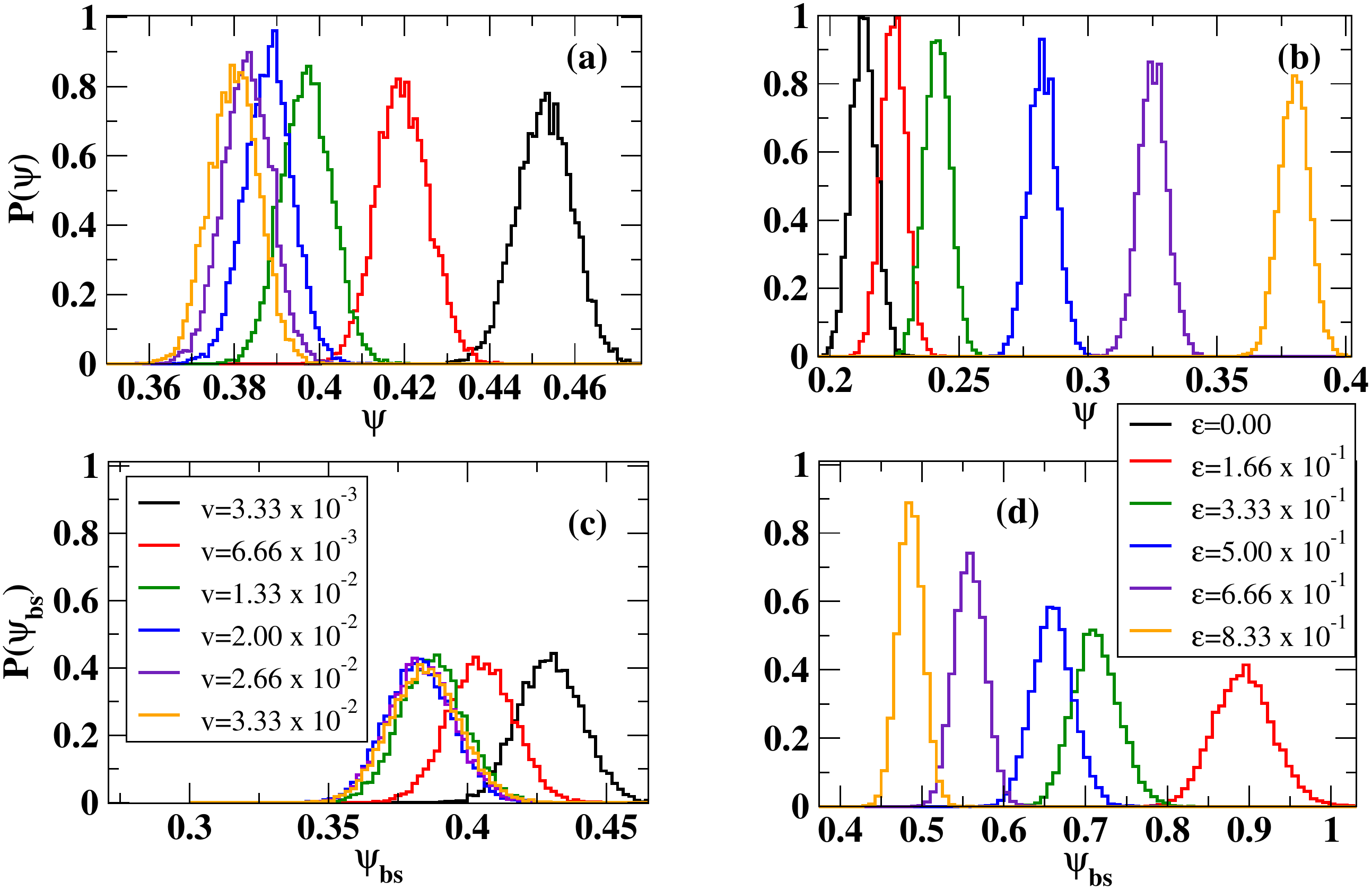} }
\caption{ (Color online) (a) Probability distribution function, $P(\psi)$ vs $\psi$ for different actyivity ($v$) and fixed polydispersity index, $\epsilon=8.33 \times 10^{-1}$. (b) $P(\psi)$ vs $\psi$ for different $\epsilon$ and fixed activity, $v=3.33 \times 10^{-2}$. (c) $P(\psi_{bs})$ vs $\psi_{bs}$ for different actyivity ($v$) and fixed polydispersity index, $\epsilon=8.33 \times 10^{-1}$. (d)  $P(\psi_{bs})$ vs $\psi_{bs}$ for different $\epsilon$ and fixed activity, $v=3.33 \times 10^{-2}$.} 
\label{fig: 407}
\end{figure} 

\subsection{Mobility order parameter}

Now we explain the enhanced dynamics due to polydispersity. First, we analyze the effect of activity on the system's dynamics, and then we study the impact of polydispersity. In a self-driven system,  particles do not stay static for a long time; instead, they keep moving throughout the system. {The crowding of the environment makes the particles collide among themselves during motion. Therefore, their instantaneous speed is not the same, but some move faster and slower.} We defined $rattlers$ in the system based on the crowding in the neighborhood of a particle.  {We call a particles $i$ is a neighbour of particles $j$ if $(\vert{\bf r_i - r_j}\vert) \le 2^{1/6}(\sigma_i + \sigma_j)$. A particle with two or less immediate neighbor (s) is called a `rattler', and hence {it is more mobile until it loses the tag}. In the $d$-dimensions, a particle can be a rattler if it has less than $d+1$ neighbour(s)}. In fig. \ref{fig: 406}(a-b), we show the snapshots from the simulation for two different polydispersities, where the particles that are rattlers are in blue, whereas those are non-rattler are in red. {We see that the rattlers prefer to be at the boundaries of the clusters of non-rattlers}. We calculate the mobility order parameter (MOP) $\psi(t)$ defined as $\psi(t)=\frac{N_r(t)}{N}$, where $N$ is the number of particles in the system and $N_r$ is the number of rattlers. Hence, $\psi \in (0,1)$,  $\psi = 1$ means that all the particles in the system are rattlers and vice versa.  This tells us that the higher the MOP value, the system will be more dynamical. Further, we refine $\psi_{bs}(t)= \frac{N_{rb}(t)}{N_{rs}(t)}$, where $N_{rb}$ and $N_{rs}$ are  the number of those rattlers whose radius is  bigger and smaller  than the mean radius $R_0$ resepctively. 
In fig. \ref{fig: 407}(a-b) we plot the probability distribution functions  $P(\psi)$ vs. $\psi$; and in  \ref{fig: 407}(c-d) we plot $P(\psi_{bs})$ vs. $\psi_{bs}$ for $\phi=0.85$ for different system parameters.  In fig. \ref{fig: 407}(a), the peaks of $P(\psi)$ shift towards smaller values of $\psi$ as we increase the activity. This suggests that the number of rattlers in the system decreases with increased activity $v$, and MSD should also decrease. But we see the opposite, because for a non-zero PDI, peak of $P(\psi_{bs})$ also shifted towards smaller values (see fig. \ref{fig: 407}(c) ). This implies that $N_{rs}$ is higher than the   $N_{rb}$, and we  know that smaller particles have higher activity (motility) (since $v_i \propto \frac{v_0}{R_i \mu \kappa}$ ) and hence, for a fixed activity $v$, the contribution of smaller particles (when $R_i \leq R_0$) to the MSD is higher than that for bigger ones (i.e. $R_i > R_0$). Eventually, we see an increase in the MSD, hence, higher $D_{eff}$ for a higher value of self-propulsion speed.  \\
Similarly, we can explain the increase in  $D_{eff}$ if we increase $\epsilon$. Fig. \ref{fig: 407}(b) shows that the peak of $P(\psi)$ shifts towards higher values and that of 
$P(\psi_{bs})$ towards left. This implies that with an increase in PDI, rattlers are increasing, but small rattlers increase more, which have much higher motility than bigger ones.  Hence the $D_{eff}$ increases with an increase in PDI.

\begin{figure}      
\centerline{\includegraphics[width=\linewidth]{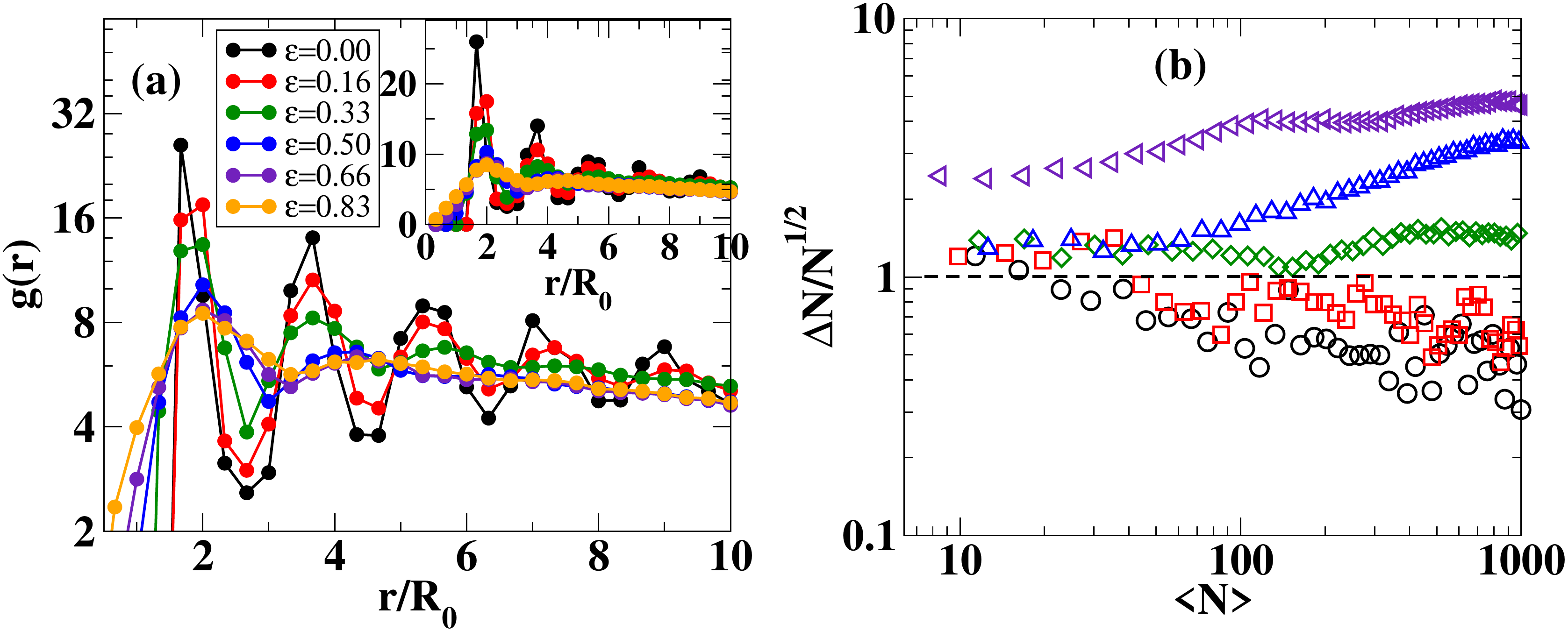} }
\caption{(Color online) (a) Plot for the radial distribution function $g(r)$ for different PDI, $\epsilon$ on semi-log scale (main), inset: plot on linear scale. We keep the activity fixed in this plot. (b) Number Fluctuation plot for $(v,\epsilon)= (3.33 \times 10^{-3}, 1.66 \times 10^{-1})$ circles: solid-jammed phase,  $(3.33 \times 10^{-3}, 6.66 \times 10^{-1})$ squares: liquid-jammed phase,  $(2.00 \times 10^{-2}, 1.66 \times 10^{-1})$ diamonds: MIPS-liquid phase, $(3.33 \times 10^{-2}, 0.0)$ triangle up: MIPS-liquid phase,  $(3.33 \times 10^{-2}, 8.33 \times 10^{-1})$ triangle left: pure liquid phase. The dashed line corresponds to $\Delta N/\sqrt{N}=1$. Both the plots are for $\phi=0.85$ and the system size $L=40$.} 
\label{fig: 408}
\end{figure}

\subsection{ Phase diagram} 
In the previous paragraph, we discussed the effect of polydispersity on the system dynamics, where we have calculated the steady-state diffusion coefficient of the system. We characterize the different phases based on the value of diffusivity, radial distribution function, and number fluctuation. First, to understand the structure of  the particles' cluster we calculate the radial distribution function (RDF) $g({r})$. Where, $g(r)$  is a measure of the probability of finding a particle at ${\bf r_2}$ given a particle at ${\bf r_1}$ ; $r= |{\bf r}_1- {\bf r}_2|$. In two dimensions $\langle n \rangle g(r) d^2{\bf r}$ gives the number of particles in $d^2{\bf r}$, where $<n>$ is the mean number of particle in unit area. 
We plot $g(r)$ vs. normalise radial distance $\frac{r}{R_0}$ in fig. \ref{fig: 408}(a), and see that with an increase in PDI, not only the height of the peak of $g(r)$ decreases, but also the distribution loses its periodicity since the number of distinct peaks ($m$) reduces. This means the structure of the distribution of the particles in the system shifted to the less ordered liquid-like structure, for bigger $\epsilon$, from the more ordered solid one, or $\epsilon = 0$. Also, if there are at least {\em three} peaks in the $g(r)$ vs. $r/R_0$ plot, this represents a near to hexagonal closed pack (HCP) structure (fig. \ref{fig: 406}a), and we call it a solid-like structure, whereas when it has less than three peaks, we call it as a liquid-like structure (far away from HCP) (Fig. \ref{fig: 406}b).

\begin{figure*}      
\includegraphics[width=\linewidth]{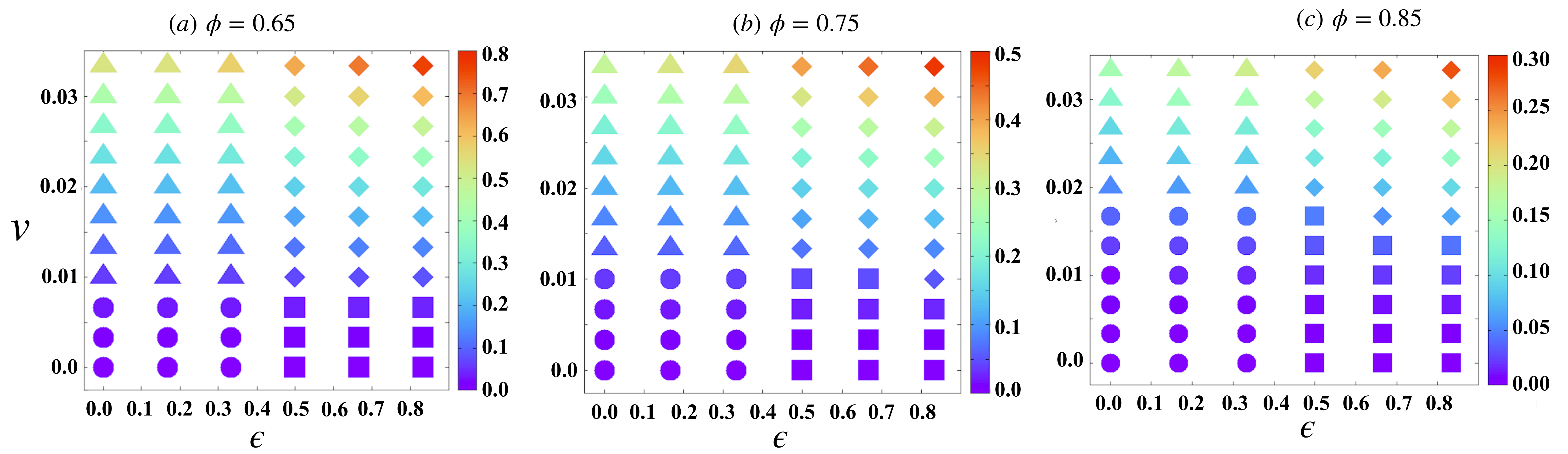} 
\caption{ (Color online) Phase diagram for $\phi=0.65$ (a),  $\phi=0.75$ (b) and $\phi=0.85$ (c): circles, squares, triangles and diamonds  represents the solid-jammed phase, liquid-jammed phase, MIPS-liquid phase  and $pure \ liquid$ phase, respectively. Color bar shows the value of $D_{eff}$ for a given ($\epsilon,v$).}
\label{fig: 409}
\end{figure*}

Second, we calculate the number fluctuation, defined as $\Delta N=\sqrt{<N^2>-<N>^2}$, where $\Delta N$ is the standard deviation in the number of particles in different size subcells {and $\langle .. \rangle$ is the the represents the average over many snapshots}. Further, $\Delta N \sim <N>^{\gamma}$, it has been found that $\gamma > 0.5$ for the ABP undergoing MIPS [\cite{FilyPRL2012,  RednerPRL2013, StenhammarPRL2013, TailleurPRL2008, GonnellaIJMPC2014, SumaEPL2014, BerthierPRL2014, WittkowskiNatComm2014}].  It suggests a large number-fluctuation in the active system undergoing dynamical phase separation.  Whereas in an equilibrium system,  $\gamma \leq 0.5$. Further, the systems is in the frozen or jammed state when $\gamma <0.5$ [\cite{HenkesPRE2011}].   In fig. \ref{fig: 408}(b), we show the plot of $\Delta N/\sqrt{N} $ vs. $<N>$ for some chosen sets of $(v,\epsilon)$, and observe that for small activity, $v \ (=3.33 \times 10^{-3}$ in the plot), system shows small number fluctuation with the curve is below the horizontal line, {which is the reference line for an equilibrium system}, and giant number fluctuation for higher activity, $v$ ($=2.0 \times 10^{-2}$  and $3.33 \times 10^{-2}$ in the plot), where the curve is above the horizontal line. Therefore we characterize the different phases in the system based on the values of $D_{eff}$, number of distinct peaks in RDF and $\gamma$ for a given set of  $( v,  \epsilon)$.  Now, we explicitly discuss the different phases in the system and show the phase diagram in fig. \ref{fig: 409}. 

{\em Jammed phases:} 
We call the system in the jammed state when $D_{eff} \leq 0.05$ and the number fluctuation exponent, $\gamma < 0.5$ for small activity. Which means that the particles in the system are almost stationary or {\em jammed}.  Further, we call it the solid-jammed state, when RDF have three or more than three peaks observed for $\epsilon \leq 3.33 \times 10^{-1}$ and the liquid-jammed phase when RDF have less than three peaks observed for $\epsilon \geq 0.5$. 

{\em Liquid Phases:} We call the system in the liquid phase when $D_{eff} \geq 0.25$ and the number fluctuation exponent, $\gamma > 0.5$ for high activity. This implies that the particles in the system are highly motile for high activity and behave like the free-flowing liquid. Further, we call it motility induced phase separation (MIPS)-liquid phase since the particles move collectively forming closed packed structure, which is observed for $\epsilon \leq 3.33 \times 10^{-1}$ and  $v \geq 0.01$. The closed packed structure is evident from fig. \ref{fig: 408}(a), where  RDF has three or more than three peaks. For large activity, enhanced motion of ABPs lead to faster accumulation near a cluster as shown in [\cite{ButtinoniPRL2013}].  
This phase is analogous the the MIPS phase reported for a mono-disperse active Brownian particles [\cite{FilyPRL2012, RednerPRL2013, StenhammarPRL2013, TailleurPRL2008, GonnellaIJMPC2014, SumaEPL2014, BerthierPRL2014, WittkowskiNatComm2014}]. We call the system in a {\em pure liquid} phase when RDF have less than three peaks and the activity $v \geq 0.01$. In this case  the particles in a dense cluster do not form an ordered pattern as it is evident from the plot of RDF $g(r)$, fig. \ref{fig: 408}(a). {We observe the enhanced diffusivity essentially in the pure liquid phase, observed for large PDI that introduces a large number of small ABPs with high motility responsible for enhanced diffusion, shown in the $D_{eff}$ vs. $\epsilon$ plot in fig. \ref{fig: 403}. }

In fig. \ref{fig: 409}, we show the phase diagram for different packing fractions. Different symbols imply the type of the phase for the given parameter set, and the color bar shows the value of $D_{eff}$ for the same. We also observe that for $\phi=0.65 \ and \ 0.75$, system show jammed phase for $v \leq 1.0 \times 10^{-2}$ whereas it is jammed for $v \leq 1.33 \times 10^{-2}$ for $\phi = 0.85$. This shift is due to the particles' high packing density, which makes the system highly crowded; hence, it needs higher activity to be in the liquid phase. \\
We find that the phases in the system are independent of the size of the system. We confirm this by plotting the $D_{eff}$ vs. $\epsilon$ for different activity and the phase diagram for $L=30$ and $\phi=0.85$ (see the Sec. \ref{appendix400}) and find an identical phase diagram to that of fig. \ref{fig: 409} (c).\\

\section{System size independence}
\label{appendix400}
\begin{figure*}      
\centerline{\includegraphics[width=\linewidth]{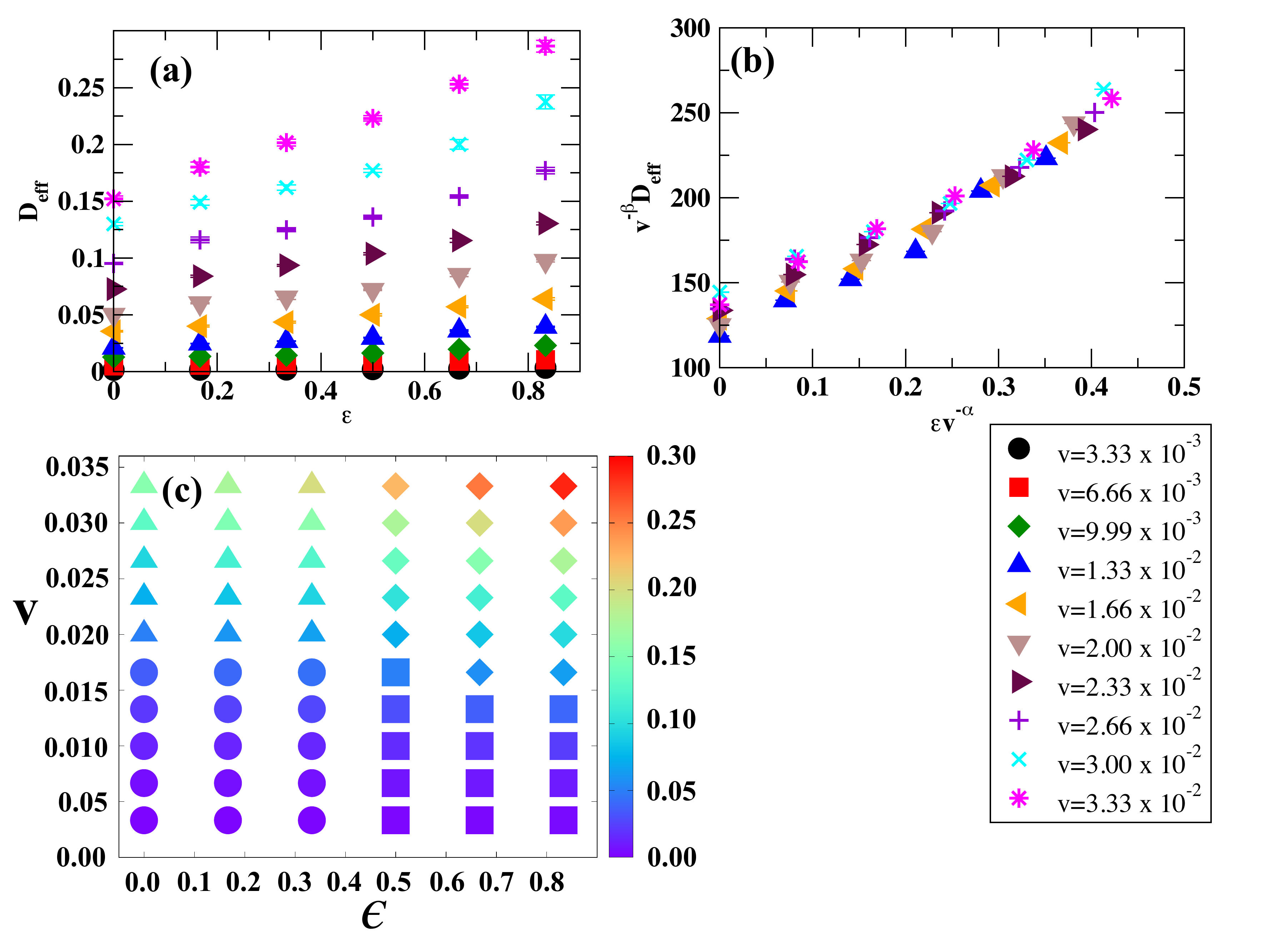} }
\caption{ (Color online) (a) Effective diffusion coefficient $D_{eff}$ vs. polydispersity $\epsilon$ for different $v$. (b) Scaled diffusivity, $D_{eff}/v_0^{\beta}$ vs. scaled PDI, $\epsilon v^{-\alpha}$, where $\alpha = -0.2$ and $\beta = 2.0$. Error bars are of the size of the symbols. (c) Phase diagram: circles, squares, triangles and diamonds  represents the solid-jammed phase, liquid-jammed phase, MIPS-liquid phase  and $pure \ liquid$ phase, respectively. Color bar shows the value of $D_{eff}$ for a given ($\epsilon,v$). All the data are generated for $\phi=0.85$ and system size $L=30$.}
\label{fig: 410}
\end{figure*}

This section shows the data for bigger system size, $L=30$, and packing density $\phi=0.85$. We see that the system's response is almost identical to what is shown for a relatively smaller system size ($L=20$) in the main text. Fig.  \ref{fig: 410}(a-b) shows the plot for effective diffusion coefficient $D_{eff}$ vs $\epsilon$ and fig. \ref{fig: 410}(c) shows the phase diagram in the plane of PDI and the activity. In the $D_{eff}$ vs. $\epsilon$ plot, we observe that the values of $D_{eff}$ is almost the same with minimal changes (within the error bars) for the chosen set of system parameters. This confirms our claim that the impact of polydispersity in a system of active Brownian particles is independent of the size of the system.

\section{Discussion}
\label{discussion400}
{We study the dynamics and the phases of self-propelled disk-shaped particles of different sizes with soft repulsive potential in two dimensions. Properties of the system are characterized for different $activity, \ (v)$, which is controlled by the self-propulsion speed of the particles, and the polydispersity index, $\epsilon$, which is the width of the uniform distribution of the particle's radius. We use over-damped Langevin's dynamics to study the particles' motion. 
We observe enhanced dynamics for large size diversity among the particles.   We calculate the steady-state diffusion coefficient $D_{eff}$ and for high activity 
$v$, it follows a scaling relation $D_{eff} \sim D_0 v^{\beta}f(\epsilon v^{-\alpha})$,  $\alpha=-0.2$  and  $\beta \simeq 2.0$.
The mobility order parameter, $\psi$ and $\psi_{bs}$, explains the enhanced dynamics for a non-zero polydispersity. We find that the dynamics of smaller particles, for large polydispersity, lead to enhanced diffusivity. We find system exhibits {\em four} distinct phases. The system is in solid jammed and liquid jammed phase for small and large PDI, for small activity. Jammed phase characterized by small $D_{eff}$. Whereas for larger activity, it forms MIPS-liquid for small PDI, when $D_{eff}$ is moderate and results match with previous MIPS in ABP. And for large PDI, we find enhanced diffusivity and no periodic structure, and the system is defined as {\em pure liquid} phase. {Further, the enhanced diffusivity observed for $pure \ liquid$ phase}. The number fluctuation is larger and smaller than the equilibrium limit in the liquid and jammed phases. We study the system for three different packing densities of the particles and observe almost the same trend.  \\
{One can also get enhanced diffusion by putting the variable speed with identical size particles similar to the work in [\cite{SinghPhysicaA2020}]. Still, in experiments,  designing the polydispersity in speed is much more challenging compared to the polydispersity in the size.}\\
Our analysis can help understand the behavior of cells of various sizes in a tissue,  artificial self-driven granular particles, or living organisms of different sizes in a dense environment. Also, one can design a similar system and obtain the results we have brought in this work.}

\bibliographystyle{apsrev4-1}
\bibliography{references.bib}

\end{document}